\pdfoutput=1
\documentclass[a4paper]{article}

\usepackage[ninepoint]{itgspeech2021}    
\usepackage{times}            
\usepackage[english]{babel}   
\usepackage[utf8]{inputenc}
\usepackage[T1]{fontenc}      
\usepackage[sort&compress,numbers]{natbib}	
\usepackage{amsmath,amssymb}
\usepackage{graphicx}
\usepackage[colorlinks=false,pdfborder={0 0 0}]{hyperref}
\usepackage{units}

\usepackage{balance}

\DeclareUnicodeCharacter{2212}{-}
\usepackage{tikz} 

\usepackage{import}
\usepackage{pgfplots}
\usepackage{pgf}
\usepackage[exponent-product=times]{siunitx}

\usetikzlibrary{positioning}
\usetikzlibrary{arrows}
\usetikzlibrary{calc}

\usepackage{amsmath}
\usepackage{amssymb}
\usepackage{bbm}
\usepackage{physics}
\usepackage{siunitx}
\usepackage{trfsigns}

\usepackage{tabularx, etoolbox, booktabs}
\usepackage{multirow} 
\newcommand{\rarray}[1]{\renewcommand{\arraystretch}{#1}} 
\newcolumntype{H}{>{\setbox0=\hbox\bgroup}c<{\egroup}@{}}  

\usepackage[capitalise]{cleveref}

\DeclareMathOperator*{\argmin}{argmin}

\newcommand{\vect}[1]{\ensuremath{\boldsymbol{\mathbf{#1}}}}

\renewcommand{\H}{^\mathrm{H}}
\newcommand{\T}{^\mathrm{T}}
\newcommand{\inv}{^{-1}}

\newcommand{\conj}{^{*}}

\newcommand{\y}{\vect{y}}
\newcommand{\x}{\vect{x}}

\renewcommand{\d}{\vect{d}}

\newcommand{\dhat}{\hat{\vect{d}}}
\newcommand{\W}{\vect{W}}
\newcommand{\w}{\vect{w}}
\renewcommand{\v}{\vect{v}}
\newcommand{\U}{\vect{U}}
\newcommand{\J}{\vect{J}}
\newcommand{\Sig}{\vect{\Sigma}}
\newcommand{\E}{\vect{E}}
\newcommand{\R}{\vect{R}}
\renewcommand{\e}{\vect{e}}
\newcommand{\I}{\vect{I}}
\newcommand{\ytilde}{\widetilde{\mathbf{y}}}
\newcommand{\B}{\vect{B}}

\newcommand{\Wtilde}{\widetilde{\vect{W}}}

\usepackage{mathtools}

\usepackage{algorithm}
\usepackage{pseudocode}
\usepackage{algorithmicx}
\usepackage{algpseudocode}

\algtext*{EndWhile}
\algtext*{EndIf}
\algtext*{EndFor}

\usepackage{xspace}
\def\wrt{w.r.t.\@\xspace}
\def\ie{i.e.\@\xspace}
\def\eg{e.g.\@\xspace}

\usepackage{multirow}

\usepackage{amssymb}
\usepackage{pifont}

\newcommand{\cmark}{\ding{51}}%
\newcommand{\xmark}{\ding{55}}%

\sloppypar

\usepackage[acronym]{glossaries}
\glsdisablehyper
\newacronym{ASR}{ASR}{Automatic Speech Recognition}
\newacronym{BSS}{BSS}{Blind Source Separation}
\newacronym{cACG}{cACG}{complex Angular Central Gaussian}
\newacronym{cACGMM}{cACGMM}{complex Angular Central Gaussian Mixture Model}
\newacronym{EM}{EM}{Expectation-Maximization}
\newacronym{HEAD}{HEAD}{Hybrid Exact-Approximate Joint Diagonalization}
\newacronym{IVA}{IVA}{Independent Vector Analysis}
\newacronym{ICA}{ICA}{Independent Component Analysis}
\newacronym{ILD}{ILD}{Inter-channel Level Difference}
\newacronym{IPD}{IPD}{Inter-channel Phase Difference}
\newacronym{MVDR}{MVDR}{Minimum Variance Distortionless Response}
\newacronym{NN}{NN}{Neural Network}
\newacronym{OverIVA}{OverIVA}{Overdetermined IVA}
\newacronym{PCA}{PCA}{Principal Component Analysis}
\newacronym{RIR}{RIR}{Room Impulse Response}
\newacronym{RTF}{RTF}{Relative Transfer Function}
\newacronym{SDR}{SDR}{Signal-to-Distortion Ratio}
\newacronym{SNR}{SNR}{Signal-to-Noise Ratio}
\newacronym{SMM}{SMM}{Spatial Mixture Model}
\newacronym{SMS-WSJ}{SMS-WSJ}{Spatialized  Multi-Speaker  WallStreet Journal}
\newacronym{STFT}{STFT}{Short-Time Fouier Transformation}
\newacronym{TF}{TF}{time-frequency}
\newacronym{Tv-G}{Tv-G}{time-varying Gaussian}
\newacronym{WER}{WER}{Word Error Rate}
\newacronym{WPE}{WPE}{Weighted Prediction Error}

\title{A Comparison and Combination of Unsupervised Blind Source Separation Techniques}

\author{Christoph Boeddeker, Frederik Rautenberg, Reinhold Haeb-Umbach}

\address{Paderborn University, Department of Communications Engineering, Paderborn, Germany\\
  Email: \texttt{\{boeddeker,haeb\}@nt.upb.de, frra@campus.upb.de}}

\DeclareMathAlphabet{\mathcal}{OMS}{cmsy}{m}{n} 

\makeatletter
\renewcommand\section{\@startsection {section}{1}{\z@}%
  {-1.5ex \@plus -0.6ex \@minus -0.1ex}{6pt \@plus.1ex}{\normalfont\Large\bfseries}}
\renewcommand\subsection{\@startsection{subsection}{2}{\z@}%
  {-1.4ex\@plus -0.5ex \@minus -0.1ex}{4pt \@plus 0.1ex}{\normalfont\large\bfseries}}
\renewcommand\subsubsection{\@startsection{subsubsection}{3}{\z@}%
  {-1.25ex\@plus -0.5ex \@minus -0.1ex}{3pt \@plus 0.1ex}{\normalfont\normalsize\bfseries}}
\makeatother
\begin{document}

\maketitle
\begin{abstract}
Unsupervised blind source separation methods do not require a training phase and thus cannot suffer from a train-test mismatch, which is a common concern in neural network based source separation.
The unsupervised techniques can be categorized in two classes, those building upon the sparsity of speech in the Short-Time Fourier transform domain and those exploiting non-Gaussianity or non-stationarity of the source signals.
In this contribution, spatial mixture models which fall in the first category and independent vector analysis (IVA) as a representative of the second category are compared \wrt their separation performance and the performance of a downstream speech recognizer on a reverberant dataset of reasonable size.
Furthermore, we introduce a serial concatenation of the two, where the result of the mixture model serves as initialization of IVA, which achieves significantly better WER performance than each algorithm individually and even approaches the performance of a much more complex neural network based technique. 
%
\end{abstract}
\section{Introduction}
%
%
In the field of speech signal processing, \gls{BSS} is concerned with separating a mixture of speech signals into the contributions of the individual speakers.
Formulating \gls{BSS} as a supervised learning problem, neural networks have excelled in this task, even if the mixture signal is recorded by a single microphone only.
However, their performance heavily depends on the absence of a mismatch between the, typically artificially mixed, training data and the test scenario.
Furthermore, neural network based single channel source separation as of today breaks down in the presence of reverberation.


On the other hand, signal processing based approaches, such as \glspl{SMM} \cite{TranVu2010SMM, itoComplexAngularCentral2016} and the class of \gls{ICA} \cite{Amari1996ICA, kim2006independent,IVA_original_2, sawada2019review, Hyvarinen2000ICA} based algorithms, are unsupervised methods, that do not require a training phase and thus cannot suffer from a train-test mismatch. 
These are multi-channel techniques and  good separation performance has been reported even if the input signals are reverberated. 

\Glspl{SMM} and \gls{ICA} based techniques rely on quite different modeling assumptions.
The core assumption of \gls{SMM} based source separation is the sparsity in the \gls{STFT} domain: at any \gls{TF} bin at most one of the sources is active, while the contribution of the others is negligibly small.
Based on this, the posterior probability of source activities for each \gls{TF} bin and the parameters of the statistical model of the multi-channel observations can be estimated with the \gls{EM} algorithm.
Typical models are the Watson \cite{TranVu2010SMM} and the \gls{cACG} \cite{itoComplexAngularCentral2016} mixture models.
The actual separation is performed with the source activity posteriors which are interpreted as masks: A source signal can be retrieved either by simply multiplying the \gls{STFT} representation of the mixture signal by the speaker's mask or by beamforming.
In the latter case, the mask is employed to compute source-specific spatial covariance matrices, from which the beamformer coefficients of common beamformers can be obtained \cite{Erdogan2016MVDR}.
Note that the number of sources $K$ can be smaller, equal or larger than the number of microphones $M$. The only requirement is that at least two microphones are available ($M \ge 2$).
It should also be mentioned that, with \glspl{SMM}, each frequency is usually treated independently, which incurs a frequency permutation problem: the order of speakers in each frequency is undefined and needs to be aligned by a so-called permutation solver, \eg, \cite{Sawada2010PermSolver}, before transformation to the time domain.

\Gls{IVA}, on the other hand, does not employ a sparsity assumption. Based on \gls{ICA}, the core assumption is that sources are either non-Gaussian, non-stationary or non-white \cite{Buchner2004trinicon}, or even non-proper \cite{Loesch2012ICAnonProper}. In case of speech considered here, non-Gaussianity and non-stationarity are the most common assumptions.
Most \gls{IVA} algorithms assume a determined case, \ie the number of speakers $K$ is equal to the number of observations $M$. This is because \gls{IVA} assumes the observation is obtained with an invertible mixing matrix from the source signals.
Source separation is thus obtained by reverting this mixing process. However, an extension to the overdetermined case where  $M \geq K$ can be obtained, e.g., by dimension reduction through \gls{PCA} \cite{Asano2000PCA_VSS}, by introducing dummy sources \cite{scheibler2019independent}
or by simply discarding $M-K$ microphone channels.
Note that IVA treats all frequencies jointly and thus does not require a frequency permutation solver.

Surprisingly, no direct comparison of \glspl{SMM} and \gls{IVA} on a reasonably sized database has been done so far, at least to the best of our knowledge.
This paper is meant to fill this gap. Here we use the \gls{SMS-WSJ} dataset proposed by \cite{drudeSMSWSJDatabasePerformance2019}.
This dataset offers up to $M=6$ observations which contain mixed signals from two speakers.
Additionally, the observations are reverberated and contain mild microphone noise.
As evaluation metric we use the \gls{SDR} \cite{vincentPerformanceMeasurementBlind2006} and the \gls{WER} of a downstream speech recognizer.

Since \glspl{SMM} and \gls{IVA} rely on quite different modeling assumptions they may have complementary weaknesses and strengths, which makes a combination of the two an attractive option.
Here, we suggest a serial concatenation, where the output of the \gls{SMM} serves as an initialization of \gls{IVA}. With this the \gls{WER} performance can be improved by more than \SI{10}{\percent} relative and approaches the best \gls{WER} reported on this dataset, which has been obtained by a much more complex multi-channel neural-network based source separator.

The rest of this paper is organized as follows, \cref{Sec:basic} describes the model and the two algorithms and the way we initialize the \gls{IVA} algorithm.
The setup of the experiments and the discussion of the results are in \cref{Sec:Experiments}.
\Cref{Sec:Conclusion} summarize this paper.

\section{Notation}
In this paper we use the following notation, small characters $x$, bold characters $\vect{x}$ and bold capital characters $\vect{X}$ defines scalars, column vectors and matrices, respectively.
$\vect{e}_k$ defines a unit vector, which is the $k$'th column of the identity matrix $\vect{I}_K \in \mathbb{R}^{K \times K}$.
The superscript $\mathrm{T}$ and $\mathrm{H}$ denote the matrix transpose and the conjugate transpose of a matrix, respectively.
All signals in this paper are in the \gls{STFT} domain, with $t$ and $f$ being the time frame and the frequency bin index, respectively.
\section{Model} 
\label{Sec:basic}
Let $\vect{s}_{f,t} = \begin{bmatrix}
 s_{f,t,1} & \hdots & s_{f,t,K}    
\end{bmatrix}\T \in \mathbb{C}^{K}$ be the vector of $K$ source signals at frequency bin index $f$ and time frame $t$. 
The ``image'' of this mixture at the  $M$ microphones is given by 
\begin{equation}
    \x_{f,t,k} \approx \sum_{\mathclap{\tau=0}}^{\mathclap {\infty}} \vect{a}_{f, \tau, k} s_{f, t-\tau, k}
\end{equation}
where $\vect{x}_{f,t,k} \in \mathbb{C}^{M}$ and where $\vect{a}_{f, \tau, k} \in \mathbb{C}^{M}$ is the vector of relative transfer functions from the source $k$ to the microphones.

The microphone signals further contain noise 
resulting in the observation vector $\vect{y}_{f,t} \in \mathbb{C}^{M}$:
%
\begin{align}
    \y_{f, t} &= \sum_{k} \x_{f, t,k} + \tilde{\vect{n}}_{f, t} \label{eq:model:obs1} \\
    &= \sum_{k}\vect{h}_{f, k} d_{f, t, k} +
    \vect{n}_{f, t}   \label{eq:model:approx_1}
    = \vect{H}_{f} \d_{f, t} + \vect{n}_{f, t} . 
\end{align}
Here, we introduced the early part of the reverberated signal $d_{f, t, k}$ and $\vect{h}_{f, k}$ represents the early part of the \gls{RTF}, while the distortion $\vect{n}_{f, t}$  captures both the noise $\tilde{\vect{n}}_{f, t}$ and the late reverberation \cite{Haeb2020ASR}.
In \cref{eq:model:approx_1}, $\vect{H}_{f} \in \mathbb{C}^{M \times K}$ and $\d_{f, t} \in \mathbb{C}^{K}$ collect the early part of the \glspl{RTF} and desired signals, respectively.



In the following, we describe two separation algorithms: \gls{SMM} with source extraction by beamforming, and \gls{IVA}.
Both conduct a linear estimation, but the core assumptions are quite different as we will see.


%
\subsection{Spatial Mixture Model}
\glspl{SMM} rely on the sparsity assumption of speech in the \gls{STFT} domain: 
At most one source is active at any time frequency bin. Let $z_{f,t} \in \{0,1, \ldots , K\}$ be a hidden variable, where $z_{f,t}=k$ indicates that source $k$ is dominant in \gls{TF} bin $(f,t)$, and $k=0$ is meant to indicated absence of a speech source.
Then sparsity implies that \cref{eq:model:approx_1} can be rewritten as
\begin{align}
    \y_{f, t} &= \begin{cases}
    \vect{h}_{f, z_{f,t}} d_{f, t, z_{f,t}}  &  z_{f,t} \in \{1, \ldots K\} \\
    \vect{n}_{f, t} &  z_{f,t} = 0.
    \end{cases}
\end{align}
With this assumption, a mixture model is appropriate to represent the distribution of $\y_{f, t}$ \cite{TranVu2010SMM}:
\begin{equation}
	p(\ytilde_{f,t}; \pi_{f,k}, \B_{f,k}) = \sum_{k} \pi_{t,k} \cdot \mathcal{A}\left(\ytilde_{f,t}; \B_{f,k} \right) \, , 
\end{equation}
where $\mathcal{A}(\cdot)$ is the component distribution, for which we used the \gls{cACG}  \cite{itoComplexAngularCentral2016} in the following experiments.
The input for the \glspl{SMM} is the observation normalised to unit length
\begin{equation}
	\ytilde_{f,t} = \frac{\y_{f,t}}{\lVert\y_{f,t}\rVert} \in \mathbb{C}^{M}\, .
\end{equation}
The rationale for the removal of the vector length is the fact that the signal amplitude is mainly determined by the source signals, while the above model is meant to capture the spatial arrangement of the sources, represented by the orientation of $\ytilde_{f,t}$, since the spatial diversity of the sources is exploited for separation. 

\subsubsection{Parameter estimation}
Maximum likelihood estimation of the \gls{SMM} parameters is achieved by the iterative \gls{EM} algorithm \cite{itoComplexAngularCentral2016}:
\begin{align}
	\gamma_{f,t,k} &= \frac{\pi_{t,k} \cdot \mathcal{A}\left(\ytilde_{f,t}; \B_{f,k}\right) }{\sum_{\Breve{k}} \pi_{t,\Breve{k}} \cdot \mathcal{A}\left(\ytilde_{f,t}; \B_{f,\Breve{k}}\right) } \, , \label{eq:smm:estep}\\
	\B_{f,k} &= M \cdot \frac{\sum_{t} \gamma_{f,t,k} \frac{\ytilde_{f,t} \cdot \ytilde_{f,t}^{\H}}{\ytilde_{f,t}^{\H} \B_{f,k}^{-1} \ytilde_{f,t}}}{\sum_{t} \gamma_{f,t,k}} \, \\
	\pi_{t,k} &= \frac{1}{F} \sum_{f} \gamma_{f,t,k} \, .
\end{align}
Here, $\gamma_{f,t,k} = \text{Pr}(z_{f,t}=k|\tilde{\vect{y}}_{f,t})$ is the posterior probability that \gls{TF} bin $(f,t)$ is dominated by source $k$.
Further, $\B_{f,k}$ is a parameter matrix. 
Please note that the mixture weight $\pi_{t,k}$ has been chosen here to be time dependent \cite{Ito2013PermFree}.
While, in theory, the time dependent mixture weight \cite{Ito2013PermFree} avoids the frequency permutation problem, experiments showed that  more reliable estimates are obtained, if a permutation solver is nevertheless integrated in the \gls{EM} algorithm, i.e., applied after each EM step \cite{LukasPHD}.
We use an unpublished\footnote{Code is published: \url{https://github.com/fgnt/pb_bss}} similarity based permutation solver from Tran Vu \cite{TranVu2010SMM} similar to \cite{Sawada2010PermSolver} and apply it after \cref{eq:smm:estep}.

\subsubsection{Beamforming}
The \gls{SMM} does not yield an estimate of the source signals.
Signal extraction is done by interpreting the posterior probabilities $\gamma_{f,t,k}^{\mathrm{ML}}$ as a mask.
The $k$-th source can be recovered simply by multiplying the observed signal with the posterior probability.
But superior performance is achieved by employing the mask to compute spatial covariance matrices, which are then used to calculate a \gls{MVDR} beamformer \cite{soudenOptimalFrequencyDomainMultichannel2010} with an \gls{SDR} based reference channel selection \cite{Erdogan2016MVDR}.
Finally, we can obtain the estimate with
\begin{equation}
    \label{eq:mvdr}
	\hat{d}_{f,t,k} = \Breve{\w}\H_{f,k} \cdot \y_{f,t}\, ,
\end{equation}
where $\Breve{\w}\H_{f,k}$ are the beamformer coefficients.

\subsection{Independent Vector Analysis}
\gls{IVA} \cite{kim2006independent,IVA_original_2} is an extension of \gls{ICA} \cite{Amari1996ICA, Hyvarinen2000ICA}.
In \gls{ICA} we assume that the observation $\y_{f,t}$ is obtained by a linear combination of the ``source'' signals $\d_{f,t}$
\begin{equation}
    \y_{f,t} = \vect{H}_{f} \d_{f,t} \, ,
\end{equation}
where it is assumed, that the mixing matrix $\vect{H}_{f} \in \mathcal{C}^{M \times K}$ is invertible.
This implies that the number of sources $K$ is equal to the number of microphones $M$.
The goal is then to find a separation matrix $\W_{f} \in \mathcal{C}^{K \times M}$ that estimates the source signals $\d_{f,t}$
\begin{equation}
\label{Eq:separation_IVA}
    \dhat_{f,t} = \vect{W}_{f} \cdot  \y_{f,t} \, ,
\end{equation}
where the key assumption is, that the source signals are independent.

The difference between \gls{ICA} and \gls{IVA} is that \gls{ICA} assumes an independent source model for each frequency, while  in \gls{IVA} the source models are coupled between the frequencies.
Here, the coupling is given by the time-varying, however frequency-independent variance $r_{t,k}$ of the source models:
\begin{equation}
    p\left(\hat{\d}_{f,t}\right) = \prod_k \mathcal{N}\left( \hat{d}_{f,t,k} ; 0, r_{t, k} \right) \, .
\end{equation}
Source separation thus exploits the nonstationarity of the sources, rather than the non-Gaussianity.\footnote{Note, though, that when averaging over the distribution of the variances (typically a Gamma distribution), the resulting predictive distribution is student-t, i.e., is supergaussian.}

The separation matrix $\vect{W}_{f}$ is commonly estimated with the maximum likelihood approach.



To accommodate the overdetermined case, where the number of microphones exceeds the number of sources ($M > K$), we here follow the approach of  Overdetermined IVA (OverIVA) in \cite{scheibler2019independent}.
They proposed to add $M-K$ dummy estimates $\v_{f, t} = \begin{bmatrix}v_{f, t, 1} & \dots & v_{f, t, M-K}\end{bmatrix}\T$, so that no dimension reduction technique is necessary.
For the $M-K$ estimates, they proposed to use a Gaussian model with a time-invariant but frequency dependent covariance, which stands in contrast to the time-varying frequency independent variances $r_{t,k}$ used for the true sources.
In \cite{scheibler2019independent} it is shown that the dummy sources $\v_{f, t}$ can be modeled as dependent variables without negatively affecting the estimation of the sources of interest $\d_{f, t}$, while at the same time allowing to find a simpler solution. Hence we use  a full covariance matrix $\Sigma_{f}$ in:
\begin{align}
    p\left(\v_{f,t}\right) = \mathcal{N}\left( \v_{f,t} ; 0, \Sigma_{f} \right)\, .
\end{align}
Now we can define the square separation matrix
\begin{align}
    \Wtilde_f = \begin{bmatrix}
        \W_f \\
        \vect{U}_f
    \end{bmatrix} \in \mathcal{C}^{M \times M} \, ,
\end{align}
and use  complex linear random variable transformation \cite{neeserProperComplexRandom1993} for 
$\y_{f, t} = \Wtilde_f\inv \begin{bmatrix}
\dhat_{f, t}\T & \v_{f, t}\T
\end{bmatrix}\T$
to obtain the likelihood:
\begin{equation}
\begin{aligned}
	L &= \mathrlap{p(\vect{y}_{f,t}; \forall f\in\{1,\ldots , F\} \text{ and } t\in\{1,\ldots, T\} ; \vect{\theta}) } \\
	&=
	\prod_{f} |\mathrm{det}(\Wtilde_f)|^2
	&&\cdot \prod_{t}
	\mathcal{N}\left( \vect{U}_{f} \y_{f,t} ; \vect{0}, \Sigma_{f} \right) \\
	&&&\cdot \prod_{k}
	\mathcal{N}\left( \w_{f, k}\H \y_{f,t} ; 0, r_{t, k} \right) \, ,
\end{aligned}
\end{equation}
where $\w_{f, k}$ is an entry of $\W_{f} = \begin{bmatrix}
\w_{f, 1} & \dots & \w_{f, K}
\end{bmatrix}\H$ and $\vect{\theta}$ are all parameters, i.e. $\W_f$ and $r_{t, k}$ for every $f$, $t$ and $k$.

\subsubsection{Parameter estimation}
Since no  closed form solution is known, the likelihood is maximized in an iterative fashion,  similar to \cite{onoAuxiliaryFunctionBasedIndependentComponent2010, onoStableFastUpdate2011}.
Algorithm \ref{al:AlgorithmIVA} summarizes the update rules for the OverIVA parameter estimation presented by \cite{scheibler2019independent}.
The maximization regarding the separation matrix leads to the \gls{HEAD} Problem \cite{yeredorHybridExactapproximateJoint2009}, which also has no closed form solution if the number of sources is greater than two \cite{onoAuxiliaryFunctionBasedIndependentComponent2010, tv_Gaussian_tutorial, onoFastStereoIndependent2012}.
So we update the separation matrix in another iterative update scheme proposed by \cite{onoStableFastUpdate2011}.
In the algorithm we used the following notation
%
\begin{align}
	\E_1 &= \begin{bmatrix}
	\I_K & \vect{0}_{K \times M-K}
	\end{bmatrix}	
	\in \mathbb{C}^{K \times M} \, , \\ 
	\E_2 &= \begin{bmatrix}
	\vect{0}_{M-K \times K} &  \I_{M-K}
	\end{bmatrix}
	\in \mathbb{C}^{(M-K) \times M} \, .
\end{align}
%
Further,
\begin{equation}
    \Sig_{\y, f} = \frac{1}{T} \sum_t \y_{f,t} \y_{f,t}^{\H} \in \mathbb{C}^{M \times M} \, ,
\end{equation}
is the covariance matrix of the observations. 
\begin{algorithm}[!t]
    \newcommand{\alignInAlg}[1]{\mathrlap{#1} \hphantom{\R_{f,k}}}
	\caption{OverIVA inspired by \cite{scheibler2019independent}}
	\begin{algorithmic}[1]
	\State Initialize $\Wtilde_f$
	\While{not converged}
      \For{\texttt{$k \in \{1, \dots, K\}$}}
        \State $r_{t,k} \leftarrow  \frac{1}{F} \sum_{f} |\vect{w}_{f,k}^{\H}  \y_{f,t}|^2 \quad \forall \, t$
        \For{\texttt{$f \in \{1, \dots, F\}$}}
            \State $\R_{f,k} \leftarrow \frac{1}{T} \sum_{t} \frac{1}{r_{t,k}} \y_{f,t} \y_{f,t}^{\H} $
        \State $\alignInAlg{\w_{f,k}}  \leftarrow \left( \Wtilde_f \R_{f,k} \right)^{-1} \e_k$
        \State $\alignInAlg{\w_{f,k}} \leftarrow \nicefrac{\w_{f,k}}{\sqrt{\w_{f,k}^{\H} \R_{f,k} \w_{f,k}}}$
		\State update $k$'th row of $\W_f $ with $\w^{\H}_{f,k}$
        \State $\alignInAlg{\J_f} \leftarrow \left( \E_2 \Sig_{\y, f} \W^{\H}_f\right)  \left( \E_1 \Sig_{\y, f} \W^{\H}_f\right)^{-1}$
        \State $\alignInAlg{\U_f} \leftarrow \begin{bmatrix} \J_f & -\I_{M-K} \end{bmatrix} $
        \State $\alignInAlg{\Wtilde_f} \leftarrow \begin{bmatrix} \W_f \\
        \U_f
    \end{bmatrix}$
      \EndFor
      \EndFor
    \EndWhile
	\end{algorithmic}
	\label{al:AlgorithmIVA}
\end{algorithm}

After the separation has been performed with \cref{Eq:separation_IVA},
the scaling ambiguity of the sources is resolved with the minimum distortion principle \cite{MinimalDistortion}
%
\begin{equation}
    \hat{\beta}_{f, k} = \argmin_{\beta_{f, k}} \sum_t \left\lVert y_{f,t,r} - \beta_{f, k} \hat{d}_{f,t,k} \right\rVert^2 \, ,
\end{equation}
where $r$ is a reference microphone index and $\hat{\beta}_{f, k}$ a scalar to fix the scaling ambiguity.
This leads to
\begin{equation}
	\hat{d}_{f,t,k} \gets \frac{\sum_{\Breve{t}} y_{f,\Breve{t}, r} \hat{d}_{f,\Breve{t},k}\conj}{\sum_{\Breve{t}}  |\hat{d}_{f,\Breve{t},k}|^2 } \hat{d}_{f,t,k} \, .
\end{equation}
where $(\cdot)\conj$ is the complex conjugate operation.
\subsubsection{Parameter initialisation}
To start the estimation iterations for \gls{IVA} an initialisation of the separation matrix $\W_f$ is necessary.
The simplest option is to employ the identity matrix for $\Wtilde_f$. A more elaborated way is to use the eigenvectors of $\Sig_{\y, f}$ that belong to the largest eigenvalues \cite{scheibler2019independent} and then calculate $\Wtilde_f$ with the help of lines 10 to 12 from \cref{al:AlgorithmIVA}.

However, initialization of an iterative algorithm turns out often times to be critical for overall performance.
Therefore, we see here an opportunity for combining the two approaches for source separation: using the result of the \gls{SMM} to initialize \gls{IVA}.

Let $\hat{d}_{f,t,k}$ be the estimates of the separated signals obtained by the \gls{SMM}. The separation coefficients $\w_{f,k}$ of \gls{IVA} can then be initialized by solving the following least squares (LS) problem
\begin{equation}
\label{eq:ls}
	\w_{f, k} = \operatorname*{argmin}_{\w_{f, k}} \left\{ \smash{\sum_{t}}  \lVert \hat{d}_{f,t, k} - \w_{f,k}\H \y_{f,t}\rVert^2 \right\} \, ,
\end{equation}
whose solution is well-known:
\begin{equation}
\begin{aligned}
		\w_f &= \left( \smash{\sum_{\Breve{t}}} \y_{f,\Breve{t}} \y_{f,\Breve{t}}^{\H} \right)^{-1}   \sum_{t} \y_{f,t}\hat{d}_{f,t, k}\conj  \, .\label{eq:iva:leastSquares}
\end{aligned}
\end{equation}
We mentioned earlier that source extraction based on \glspl{SMM} is done by beamforming.
Therefore, it appears natural to use the beamforming vectors, such as $\w\H_{f,k}$, \cref{eq:mvdr}, directly as the initial values of the separation matrix.
Indeed, this is a valid option, since the beamformer coefficients have been derived as linear estimates that optimize some objective function, such as mean squared error or minimum variance under a distortionless constraint.
In the following we will, for reasons to be explained below, use different \gls{STFT} sizes for \gls{SMM} and \gls{IVA} based separation.
While transforming the coefficients from one \gls{STFT} size to another is in principle possible, it is easier to obtain the coefficients in the target \gls{STFT} window size by solving the above LS problem.
%
Note, in case of the same \gls{STFT} window size and shift in \gls{SMM} and \gls{IVA}, the beamformer is equal to the least squares estimate from \cref{eq:iva:leastSquares}, which can be seen, when we use \cref{eq:mvdr} in \cref{eq:ls}:
\begin{equation}
	\w_{f, k} = \operatorname*{argmin}_{\w_{f, k}} \left\{ \smash{\sum_{t}}  \lVert \Breve{\w}\H_{f,k} \cdot \y_{f,t} - \w_{f,k}\H \y_{f,t}\rVert^2 \right\} \, .
\end{equation}

\section{Experiments}
\label{Sec:Experiments}

%
For evaluating the models, we use the \gls{SMS-WSJ} dataset proposed in \cite{drudeSMSWSJDatabasePerformance2019}.
This dataset uses the audio data from the WSJ Database \cite{Paul1992WSJ}, resampled to \SI{8}{kHz} sampling rate.
Utterances are artificially reverberated by convolving with simulated \glspl{RIR} with the sound decay time $T_{\text{60}}$ sampled uniformly in the range from $200$ to $\SI{500}{ms}$, to create observations for $6$ microphones.
Two utterances are added at an average \gls{SDR} of $\SI{0}{dB}$ to create a mixture. Further,
to simulate the microphone noise, Gaussian noise with an average \gls{SNR} of \SI{25}{\decibel} is added.

The models in this paper don't require a training, so only the test set is used.
This set contains \num[group-separator={\,},group-minimum-digits=3]{1332} mixtures, \num[group-separator={\,},group-minimum-digits=3]{45144} spoken words giving rise to \num[round-mode=places,round-precision=0,]{201.3929875} minutes of test data.

As performance metrics we use the \gls{SDR} \cite{vincentPerformanceMeasurementBlind2006} and the \gls{WER}.
For the \gls{WER} calculation we used the  Kaldi \cite{Povey2011Kaldi} \gls{ASR} model for this database \cite{drudeSMSWSJDatabasePerformance2019}.

\begin{table}[b]
    \setlength{\abovecaptionskip}{0ex}
    \caption{Reference scores on SMS-WSJ. Early image is the speech source convolved with the initial part of the \gls{RIR} \cite{drudeSMSWSJDatabasePerformance2019}.}
    \label{Tab:Oracles}
	\centering
	\setlength{\tabcolsep}{5.4pt}
	\rarray{0.9}
	\footnotesize
    \begin{tabular}{l c  S[round-precision=2,round-mode=places, table-format = 1.2] S[round-precision=2,round-mode=places, table-format = 1.2]}
        \toprule 
        {} & {Size / Shift} &  {WER [\%]} &   {SDR [dB]} \\
        \midrule
        Observation & - / -         &   78.12 &  -0.009609 \\
        \midrule
        Early image   & - / -         &    7.24 &  57.097230 \\
        + linear constraint: \cref{eq:ls} & 1024 / 128 &   11.64 &  16.331512 \\
        + linear constraint: \cref{eq:ls} & 2048 / 256 &    9.29 &  20.066878 \\
        \bottomrule 
    \end{tabular}
\end{table}
\subsection{Baseline and Topline}
The first row of \cref{Tab:Oracles} shows  the \gls{SDR} and the \gls{WER} if the observations are taken as they are, i.e., without  any separation.
Furthermore, for each speaker separately, the early image signals\footnote{The early image is the speech source convolved with the initial \SI{50}{ms} of the \gls{RIR}} are used \cite{Haeb2020ASR, drudeSMSWSJDatabasePerformance2019}. This corresponds to perfect dereverberation and source separation, serving as topline for our experiments.

Next, those image signals are used  as initialization for the IVA separation matrix, i.e., replacing ${\dhat}_{f,t}$ by $\d_{f,t}$ in \cref{eq:ls}, which can serve as another indication of which performance is best possible.
Note that the performance strongly depends on the STFT window size and the STFT advance, with longer windows leading to improved performance up to $2048$.
This was to be expected because the approximation of a convolution by a simple multiplication in the STFT domain, \cref{eq:model:approx_1}, (so-called multiplicative transfer function approximation \cite{Avargel2007multiplicativeTransferFunction}) is better justified.

\begin{table}
    \setlength{\abovecaptionskip}{0ex}
	\caption{Scores for \gls{SMM} and \gls{IVA} with different \gls{STFT} parameters.}
	\label{Table:results_single_mode}
	\centering
	\setlength{\tabcolsep}{6pt}
	\rarray{0.9}
	\footnotesize
	\begin{tabular}{lc  S[round-precision=2,round-mode=places, table-format = 1.2] S[round-precision=2,round-mode=places, table-format = 1.2]}
		\toprule  
		{} & Size / Shift  &    {WER [\%]} & {SDR [dB]} \\
		\midrule
		SMM \cite{drudeSMSWSJDatabasePerformance2019} & 512 / 128 & 18.70 &  12.3 \\
		SMM & 1024 / 128 & 14.33 & 12.801043132195153 \\
		SMM & 2048 / 256 & 17.53 & 11.15374523217886 \\
		IVA & 1024 / 128 & 13.84 & 13.530299192853118\\
		IVA & 2048 / 256 & 11.80 & 13.719108047691964\\
		\bottomrule 
	\end{tabular}
\end{table}
\subsection{STFT Sizes}
\label{Sec:STFT_sizes}
In all following simulations, $M=6$ microphones are used for separation and the number of considered sources are $K=3$.\footnote{We observed better performance, if we used a separate class for the noise: $2$ speakers plus noise amount to a total of $3$ classes.}

Following up on the effect of the STFT size, 
\cref{Table:results_single_mode} shows the performance of the source separation algorithms SMM and IVA for different STFT window sizes and shifts. 
The result in the first row is the baseline result from \cite{drudeSMSWSJDatabasePerformance2019}.
It can be observed that SMM obtains its best results for an STFT size of 1024, while IVA can be further improved by increasing the size to 2048. This can be explained by the fact that SMM builds upon the sparsity of speech in STFT domain, which is lost if the STFT size is chosen too large, while IVA does not have to bother with sparsity and therefore can afford larger sizes.

Further, it is striking that IVA clearly outperforms SMM based source separation. It should also be mentioned that the SDR is not always a good indicator of WER performance. Comparing the results in the last two rows, although the SDR improves by only $\SI{0.19}{dB}$, the WER decreases by more than two percentage points! We will come back to this discrepancy below.

\begin{table}
    \setlength{\abovecaptionskip}{0ex}
	\caption{Results of the \gls{IVA} algorithm, using the mixture model as initialization}
	\label{Table:initialisierung}
	\centering
	\setlength{\tabcolsep}{7.2pt}
	\rarray{0.9}
	\footnotesize
	\begin{tabular}{cccH  S[round-precision=2,round-mode=places, table-format = 1.2] S[round-precision=2,round-mode=places, table-format = 1.2]}
		\toprule  
		Initiali- & SMM & IVA  &  time  & {WER [\%]} & {SDR [dB]} \\
		zation & Size / Shift & Size / Shift  & {domain} &   {} & {} \\
		\midrule
		\multirow{2}{*}{$\w_{f,k}^{\mathrm{BF}}$} & 1024 / 128 & 1024 / 128 & \xmark & 12.57 & 13.7197367812434\\
		 & 2048 / 256 & 2048 / 256 & \xmark & 10.84 & 13.843050404175468\\
		\midrule
		\multirow{3}{*}{$\dhat_{f,t,k}$} & 1024 / 128 & 1024 / 128 & \cmark & 12.56 & 13.714369856968073\\
		& 1024 / 128 & 2048 / 256 & \cmark & 10.69  &   13.876261325189136\\
		& 2048 / 256 & 2048 / 256 & \cmark & 10.84 & 13.854108095622134 \\
		\bottomrule 
	\end{tabular}
\end{table}
\subsection{Model Chaining}
\Cref{Table:initialisierung} displays results obtained with the initialization of the IVA separation matrix using the \gls{SMM} output.  
Using the initialization improves the results of \gls{IVA}. While the \gls{SDR} is improved by only \SI{0.15}{\decibel}, the \gls{WER} improved by more than $1$ percentage point compared to the best results in \cref{Table:results_single_mode}.  

\begin{figure}
	\centering
	\scalebox{0.3}{\input{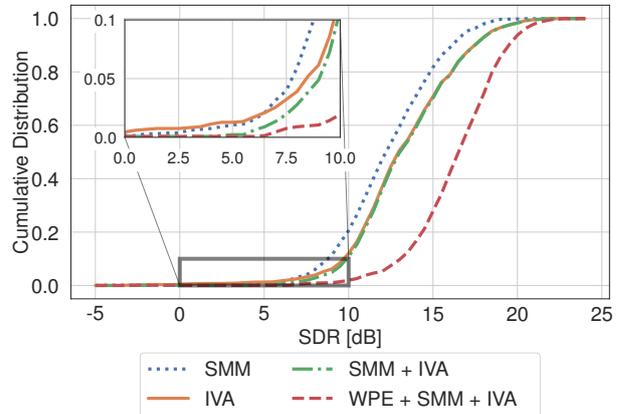}}
	\caption{Cumulative distribution of the achieved SDR values on the testset}
	\label{Fig:cumulative_sdr}
\end{figure}

In order to shed some light on the significance of the SDR on WER, \cref{Fig:cumulative_sdr} displays the cumulative distribution function  of the \gls{SDR} for different models. 
Here we can see that \gls{IVA} achieves very low \gls{SDR} values for some examples:
Without the SMM-based initialization, \SI{3.2}{\percent} of the examples separated by \gls{IVA} achieved a \gls{SDR} of $\leq 7\,$dB. 
With the SMM-based initialization, the number of poorly separated mixtures is reduced to \SI{2}{\percent}. 

Mixtures with such low separation performance will cause recognition errors in the ASR engine. Thus, reducing the percentage of poorly separated mixtures may not have a large impact on the average SDR performance, but nevertheless have significant effect on the average WER.

\subsection{Dereverberation and Comparison with State-of-the-Art}
Since the data is reverberated and since the source separation algorithms by themselves are not meant to carry out dereverberation, we experimented with \gls{WPE} \cite{Nakatani2010WPE,Drude2018NaraWPE} as a preprocessing step in front of source separation.
\gls{WPE} is a powerful dereverberation algorithm which has led to improved \gls{ASR} performance on many databases.

Here, we apply \gls{WPE} using an \gls{STFT} with a window size of $512$ and a shift of $128$. Those parameter values have been proposed in earlier studies and can be motivated by the fact that \gls{WPE} explicitly aims at extracting the direct signal and early reflections, up to $\SI{50}{ms}$ of the \gls{RIR}.
Therefore, relatively short STFT windows should be used such that the late reverberation is captured by the windows following the window containing only the direct signal and early reflections.

With \gls{WPE} preprocessing and initializing \gls{IVA} with \gls{SMM} we obtain the overall best results with $9.55\%$ \gls{WER} and $\SI{16.84}{dB}$ \gls{SDR}. 

In \cref{Table:final} we have added the  results reported in \cite{Wang2020MultimicrophoneComplexSpectral}.
They tried different combinations of \glspl{NN} and beamforming.
The best configuration consisted of first a \gls{NN}, then beamforming and then again a \gls{NN}, where both \glspl{NN} had access to all microphones and the objective was to do dereverberation and source separation.
The system which achieved the currently best \gls{WER} on SMS-WSJ  is tuned to the specific array configuration and is a huge system with $14$ million parameters.


We note that our approach proposed here is worse by about one percentage point, but achieves this result with an unsupervised approach and much less parameters.
Further, \gls{SMM} and \gls{IVA} are array geometry independent and have already proven to work with real recordings.

Since beamforming alone didn't worked well in \cite{Wang2020MultimicrophoneComplexSpectral}, 
see entry NN+BF \cite{Wang2020MultimicrophoneComplexSpectral} in \cref{Table:final},
we applied the \gls{NN} system from \cite{Boeddeker2021CISDR} to this dataset (entry NN+BF \cite{Boeddeker2021CISDR}).
When comparing to our proposal that uses a linear enhancement (i.e. beamforming), we see an advantage of ours compared to the \glspl{NN}.



%

%

%

%
 \begin{table}
    \setlength{\abovecaptionskip}{0ex}
	\caption{Results of the proposed chain and comparison with literature}
	\label{Table:final}
	\centering
	\setlength{\tabcolsep}{6pt}
	\rarray{0.9}
	\footnotesize
	\begin{tabular}{l c  S[round-precision=2,round-mode=places, table-format = 1.2] S[round-precision=2,round-mode=places, table-format = 1.2]}
		\toprule  
		& Array & {WER [\%]} & {SDR [dB]} \\
		& agnostic &  &  \\
		\midrule
		WPE + SMM & \cmark & 11.52 &  15.575920546213537\\
		WPE + IVA & \cmark & 10.91 & 16.60314576416416 \\
		WPE + SMM + IVA & \cmark  & 9.55 & 16.841328210401475\\
		\midrule
		NN + BF + NN \cite{Wang2020MultimicrophoneComplexSpectral} & \xmark  &  8.52 & {-} \\
		NN + BF \cite{Wang2020MultimicrophoneComplexSpectral} & \xmark  &  28.71 & {-} \\

		NN + BF \cite{Boeddeker2021CISDR} & \cmark  &  16.84 & 14.198593 \\
		
		\bottomrule 
	\end{tabular}
\end{table}

\section{Conclusion}
\label{Sec:Conclusion}
In this paper, two unsupervised blind source separation techniques are compared on the \gls{SMS-WSJ} dataset. 
It is shown that \gls{IVA} outperforms \gls{SMM}.
Furthermore, initializing IVA with the output of SMM further improved the WER performance of IVA by about 1 percentage point or \SI{10}{\percent} relative.
In a comparison with results from the literature, the unsupervised techniques have shown to get close to the best \glspl{NN}, without being array dependent and outperform array agnostic \gls{NN} approaches.



\small
\bibliographystyle{ieeetr}
\bibliography{example}
\balance

\end{document}